# Thermodynamic theory of highly multimoded nonlinear optical systems

Fan O. Wu, Absar U. Hassan & Demetrios N. Christodoulides *

CREOL, College of Optics and Photonics, University of Central Florida, Orlando, Florida 32816–2700, USA.

*demetri@creol.ucf.edu

Abstract

The quest for ever higher information capacities has brought about a renaissance in multimode optical waveguide systems. This resurgence of interest has recently initiated a flurry of activities in nonlinear multimode fiber optics. The sheer complexity emerging from the presence of a multitude of nonlinearly interacting modes has led not only to new opportunities in observing a host of novel optical effects that are otherwise impossible in single-mode settings, but also to new theoretical challenges in understanding their collective dynamics. In this Article, we present a consistent thermodynamical framework capable of describing in a universal fashion the exceedingly intricate behavior of such nonlinear highly multimoded photonic configurations at thermal equilibrium. By introducing pertinent extensive variables, we derive new equations of state and show that both the "internal energy" and optical power in many-mode arrangements always flow in such a way so as to satisfy the second law of thermodynamics. The laws governing isentropic processes are derived and the prospect for realizing Carnot-like cycles is also presented. In addition to shedding light on fundamental issues, our work may pave the way towards a new generation of high power multimode optical structures and could have ramifications in other many-state nonlinear systems, ranging from Bose-Einstein condensates to optomechanics.

Recent years have witnessed a strong comeback of multimode (MM) fiber technologies, largely in anticipation of high-speed communication systems that benefit from space division multiplexing[1-6]. These activities have, in turn, incited a renewed attention in the nonlinear properties of such many-mode structures[7-9] aimed at establishing new platforms for high-power fiber-based light sources[10]. During the course of this effort, a number of intriguing processes have been observed that have no counterpart whatsoever in single-mode settings. These include for example, geometric parametric instabilities[11-14], spatiotemporal mode-locking[10], efficient supercontinuum generation[12,15], and the formation of multimode solitons along with a novel class of Cherenkov dispersive wave lines[16,17], to mention a few. In the same vein, in three independent studies, a peculiar effect was found to take place in such nonlinear multimode environments whereby the optical power gradually flowed towards the lowest group of modes[12,18,19]. This beam self-cleanup mechanism, which so far remains poorly understood, seems to result from the conservative component of the Kerr nonlinearity, while having no ties to any stimulated Raman and/or Brillouin effects. Even though one can in principle address this perplexing behavior by resorting to global or MM-nonlinear wave solvers[20], it is still impossible to either predict or decipher the convoluted response of such heavily multimoded systems. For example, just setting up a multimode propagation code for individually tracking $M$ modes, will first require a computation of $M$ different dispersion curves, $M$ self-phase modulation coefficients, and asymptotically $M^2$ cross-phase modulation constants and $M^3$ four-wave mixing products[7,21] – something that is virtually unattainable, especially in systems involving thousands of modes. More importantly, such approaches lend no insight as to how this energy exchange between modes transpires or how it can be harnessed to one's advantage. Clearly of interest will be to develop an appropriate formalism capable of providing the laws that dictate the collective dynamics of such optical multimoded nonlinear configurations. To some extent, this calls for a theory akin to that of thermodynamics or statistical mechanics that are known to serve as powerful tools in understanding the macroscopic properties of various phases of matter[22,23]. More striking is the fact that such descriptions can capture the physics of these many-body systems (even when involving Avogadro-like numbers), often without delving into the underlying nature of particle-particle interactions. In this regard, non-equilibrium kinetic formulations based on optical wave turbulence theories have been put forward in order to understand such behaviors in a number of settings[24,25]. Yet, an equilibrium thermodynamic theory capable of describing such process in non-extended heavily nonlinear multimode optical structures is still lacking.

In this Article, we show that under thermal equilibrium conditions, the nonlinear evolution dynamics in conservative optical arrangements with a large but finite number of modes, can be rigorously described through a comprehensive thermodynamic formulation. These systems come with their own laws. The results derived here are universal in the sense that they apply to both continuous and discrete systems that evolve in either space or time – irrespective of the specific type of nonlinearity involved. During the process of thermalization, the total entropy always increases in such a way that the "internal energy" flows from a hotter to a colder subsystem while any exchange of optical power is driven by the difference in chemical potentials. In this respect, we derive a new set of equations of state and we cast the fundamental thermodynamic equation of entropy in terms of the extensive variables associated with the internal energy, number of modes, and optical power. Once the eigen-spectrum of a specific system is known, one can then uniquely predict its equilibrium state from these three conserved quantities. In addition, the invariants governing isentropic compressions or expansions are presented. Finally, we discuss the possibility for negative temperatures – an equilibrium regime that happens to be completely opposite to that of beam self-cleaning.

To illustrate our approach, let us consider an arbitrary nonlinear multimode optical waveguide supporting a finite number of $M$ bound states – all propagating along the axial direction $z$. In general, this configuration can be continuous in nature[21], having an elevated refractive index profile $n(x, y)$, or discrete, like for example a multicore optical fiber or waveguide array[26,27]. Each mode $i$ is associated with a particular propagation constant and an orthonormal eigenfunction $|\psi_i\rangle$ as obtained from the pertinent eigenvalue problem. The distribution of the normalized propagation eigenvalues $\varepsilon_i$ constitutes the eigen-spectrum of the system (see Supplementary information for all normalizations used in this work). In all occasions, the optical power $\mathcal{P}$ propagating in this system is conserved – as expected under continuous-wave or broad pulse excitation conditions[12,18,19]. An additional invariant, is the system's Hamiltonian which is comprised of a linear and a nonlinear component, i.e., $H = H_L + H_{NL}$ (see Supplementary). As in recent experiments, we will assume that the power levels in this nonlinear MM arrangement are relatively low and hence the Hamiltonian is heavily dominated by the linear contribution, $H \simeq H_L$. As we will see, the role of nonlinearity is to allow for a random power exchange among the various modes. To some extent, this is analogous to a 'diluted gas of particles' whose internal energy is dominated by its kinetic part while the corresponding intermolecular potential energy can be neglected – even though it is responsible for thermalization through particle collisions[22].

At this point, the conserved internal energy $U$ of this optical MM system is defined as $U = -H_L$, where the expectation value of the linear Hamiltonian operator is given by $H_L = \langle \Psi | \hat{H}_L | \Psi \rangle = \sum_{i=1}^{M} \varepsilon_i |c_i|^2$ (Supplementary information). In this latter expression, $|c_i|^2$ represent modal occupancy coefficients and are related to the total power via $\mathcal{P} = \sum_{i=1}^{M} |c_i|^2$. Note that the two *independent variables* $U$ and $\mathcal{P}$ are completely determined by initial excitation conditions, i.e., $\mathcal{P} = \sum_{i=1}^{M} |c_{i0}|^2$ and $U = -\sum_{i=1}^{M} \varepsilon_i |c_{i0}|^2$ where $c_{i0} = \langle \psi_i | \Psi_0 \rangle$ stand for the complex coefficients resulting from the projection of the input field $|\Psi_0\rangle$ on the respective modes $|\psi_i\rangle$, right at the input. While the modal occupancies $|c_i|^2$ can vary significantly during propagation as a result of weak nonlinearity – a necessary ingredient for thermalization – they are always reshuffled in a manner that $U$ and $\mathcal{P}$ remain invariant. A similar discussion holds for conservative nonlinear optical MM-cavity configurations[28,29] that evolve in time, where in this case, the eigenvalues $\varepsilon_i$ are now expressed in the frequency domain while the stored energy $E$ plays the role of $\mathcal{P}$.

In laying down our formulation, we assume that the non-integrability of the underlying nonlinear interactions enables the system to behave ergodically, thus allowing it to explore its constant energy ($U$) and power ($\mathcal{P}$) manifolds or all its accessible microstates in a fair manner – covering it uniformly with respect to the microcanonical probabilities[30]. Within this isolated microcanonical ensemble, we may then pose the following question: If the total optical power is subdivided into a very large number of indistinguishable packets (each carrying the same infinitesimal amount of power), in how many ways can one distribute them into $M$ distinct modes – subject to the constraint imposed by $U$ and $\mathcal{P}$ being constant? Given that, under these conditions the number of such packets per state is exceedingly high, maximization of entropy directly leads to a subcase of the Bose-Einstein distribution – the so-called Rayleigh-Jeans (RJ) distribution (see Methods). In this case, after thermalization, the expectation values of the mode occupancies can be obtained from $|c_i|^2 = -T/(\varepsilon_i + \mu)$, where the optical temperature $T$ and chemical potential $\mu$ result from the Lagrange multipliers associated with the constraints imposed by $U$ and $\mathcal{P}$, respectively. Here, the dimensionless quantities $T$ and $\mu$ represent intensive properties of this system and have nothing to do with the actual thermal environment the optical MM-arrangement is embedded in. Along these lines, the entropy of the system can now be obtained from Boltzmann's expression ($S = \ln W$) which leads to (see Methods)

$$S = \sum_{i=1}^{M} \ln|c_i|^2 \quad (1)$$

Interestingly, the entropy in Eq. (1) can in addition describe non-equilibrium processes and provides a basis in establishing a particular form of Boltzmann's $H$ function[31]. By adopting non-equilibrium formulations, previous studies have considered multi-wave mixing kinetics in extended nonlinear systems, ranging from $\chi^{(2)}$ materials to saturable and non-local media[24,32-36]. On the other hand, thermal equilibrium (leading to an RJ distribution) can only be reached in non-extended nonlinear optical arrangements (having a finite number of modes), as long as the underlying nonlinear mechanisms promote dynamical chaos so as to establish ergodicity. In this respect, chaos is expected to take place in the system whenever the additional invariants needed for integrability (besides the Hamiltonian and the norm) are absent. The RJ distribution can result under general conditions[37-39] (see also Methods), an aspect that was also highlighted in previous studies, even under static conditions[33,40]. We note that the nature of the equilibrium thermodynamical problem at hand is such that it requires a modal formulation, as opposed to a local description of amplitude-phase statistics at distinct sites[41-45].

From the Rayleigh-Jeans distribution and the two system invariants $U$ and $\mathcal{P}$ we can then derive the following equation of state (see Methods),

$$U - \mu\mathcal{P} = MT \quad (2)$$

which relates the two intensive variables $T$ and $\mu$ to the three extensive quantities $U$, $M$ and $\mathcal{P}$. To some extent, this equation is analogous to that of an ideal gas, e.g. $pV = Nk_BT$. Equation (2) can be used to uniquely determine both the temperature and chemical potential at thermal equilibrium, once the internal energy and power are specified at the input of a nonlinear optical system having $M$ modes (see Methods). Figure 1a shows a schematic of an optical multimode fiber involving $M \simeq 480$ modes, when excited at a normalized power $\mathcal{P} = 42.7$ and internal energy $U = 3.2 \times 10^3$, as dictated by the input power distribution among modes (Fig. 1b). For these initial conditions, our theoretical model predicts at steady-state, a temperature $T = 5.37$ and a chemical potential $\mu = -136.2$. Numerical simulations performed on this same microcanonical ensemble system are in excellent agreement with these predictions, once thermalization is attained (i.e., the entropy $S$ is maximized). Note that in this case, most of the power eventually finds its way to the lowest group of modes (Fig. 1b) – a behavior consistent with the process of beam self-cleaning observed in experiments. A similar scenario was previous considered in ref. [40,46], where the use of a finite number of modes was crucial in regularizing the UV catastrophe that could result from the RJ distribution. We next carry out simulations on a discrete (multicore) nonlinear optical waveguide array, in this case a Lieb lattice (Fig. 1c) which is known to exhibit a massive degree of degeneracy in its flat band, where $\varepsilon_i = 0$. For the $U, \mathcal{P}$ input parameters used in this example (as specified in Fig. 1d), we now theoretically expect a *negative* temperature $T = -0.76$ with a chemical potential $\mu = 6.6$. Again, the numerically obtained final power distribution among modes, after thermalization, is in full agreement with that anticipated from theory – only this time the power tends to flow towards the highest group of modes (Fig. 1d), in direct contrast to beam self-cleaning. Negative temperatures are common to systems with a finite number of modes, when the internal energy exceeds the mean value of the eigenvalue spectrum. Such configurations are known to be "hotter than hot"[47-49], i.e., the internal energy should always flow from a negative temperature region to a positive one. Similar results are also obtained for a three-dimensional array of optical cavities (CROWs) where thermalization takes place instead in time (Fig. 1e,f). In all our simulations, we make sure that the linear part of the Hamiltonian remains invariant during evolution and hence the system behaves in a quasi-linear manner for the particular power levels used. In this regime, any soliton formation (representing a phase transition) is inhibited.

We next express the entropy $S$ of a microcanonical optical system as a function of three extensive variables, $S = S(U, M, \mathcal{P})$, in a way similar to that employed in standard thermodynamics[22], where now the number of modes plays the role of volume and the optical power is analogous to the number of particles involved. As opposed to alternative formulations provided in previous studies, we here establish the fundamental thermodynamic equation $S(U, M, \mathcal{P})$ in terms of all these three conserved quantities. In fact, failure to do so, violates the very extensivity of the entropy itself. To justify this argument, let us double, for example, the input power and energy $(\mathcal{P}, U)$, as well as the number of modes $M$ in a system. From the equation of state, Eq. (2), one quickly concludes that in this case, the temperature and chemical potential remain the same as before enlarging this arrangement. We emphasize that in doubling the number of modes, the *eigen-* *spectrum distribution* should remain invariant, in other words, the "material composition" of the system should not change (see Methods). Under these conditions, each eigenvalue $\varepsilon_i$ splits into a closely spaced doublet, and hence from $S = \sum \ln|c_i|^2$, we find that $S \to 2S$. This directly implies that $S(\lambda U, \lambda M, \lambda \mathcal{P}) = \lambda S(U, M, \mathcal{P})$ – hence guaranteeing the extensivity of the entropy with respect to $(U, M, \mathcal{P})$, as required by a self-consistent thermodynamic theory. From here, one can obtain the following conjugate *intensive* variables: temperature-$T$, chemical potential-$\mu$ and pressure-$p$, via

$$\frac{1}{T} = \frac{\partial S}{\partial U}, \quad \frac{\mu}{T} = -\frac{\partial S}{\partial \mathcal{P}}, \quad \frac{p}{T} = \frac{\partial S}{\partial M} \quad (3)$$

We note that the aforementioned definitions for $1/T$ and $\mu/T$ are congruent with the equation of state Eq. (2) (see Methods). Meanwhile, it can be readily shown that $\partial S/\partial M = p/T = (S/M) - 1$. By using the fact that the entropy $S$ is a homogeneous function of $(U, M, \mathcal{P})$, the Euler equation (see Methods) leads to a second equation of state that now relates seven thermodynamic variables in this nonlinear multimode optical system, i.e.,

$$ST = U - \mu\mathcal{P} + pM \quad (4)$$

We now illustrate the versatility of the formalism developed above in predicting the thermodynamic behavior of complex, nonlinear heavily multimoded optical arrangements. Figure 2 illustrates two square lattice systems each involving $20 \times 10$ sites. The array on the left is excited with a left-hand circular polarization where the input power and energy are $\mathcal{P}_L = 41.5$ and $U_L = 15.4$. On the other hand, the lattice on the right is excited with a right-hand circular polarization with $\mathcal{P}_R = 17.9$ and $U_R = -34.9$. Numerical simulations show that these two subsystems finally reach thermal equilibrium at $(T_L, \mu_L) = (-2, 10)$ and $(T_R, \mu_R) = (0.25, -4.75)$, in full accord with theory. Eventually, they are brought in contact, thus forming a regular square lattice with 400 sites. In this new canonical-like ensemble, the total energy $U_T = U_L + U_R$ is conserved, while the respective power components $\mathcal{P}_L$ and $\mathcal{P}_R$ remain invariant. After merging, the two polarizations start to exchange energy $dU$ through cross-phase modulation (Supplementary information), where the change in the total entropy ($S_T$) is governed by the second law of thermodynamics, $dS_T = (T_L^{-1} - T_R^{-1})dU_L \geq 0$, (see Methods). This implies that the internal energy always flows from hotter to colder objects, in this case from the left circular polarization states to the right. Once the composite system is thermalized at equilibrium, the two polarizations attain the same temperature $T = 1.12$ while the chemical potentials settle to different values $\mu_L = -11$ and $\mu_R = -25$ (since no power exchange occurs), as predicted by theory and confirmed by simulations. The situation here is analogous to that of an ideal gas involving two species of particles (like oxygen and nitrogen), set at different initial temperatures and then allowed to mix in the same vessel – thus reaching the same final temperature but different chemical potentials. Figure 3 depicts another situation where two different array subsystems or "solids" having different band structures are brought into thermal contact. The rectangular lattice on the left is excited with $\hat{x}$-polarized light and is at thermal equilibrium $(T_x, \mu_x) = (0.05, -3.51)$. Meanwhile, light in the right (graphene) lattice is $\hat{y}$-polarized and at steady state reaches $(T_y, \mu_y) = (1, -14)$. The thermal contact layer in between allows the two linear polarizations to locally interact via cross-phase modulation without again exchanging power (for a possible design, see Supplementary Methods Fig. S2). Unlike the previous situation, the two polarizations are confined in their respective arrays, only exchanging energy $U$ through the thermally permeable wall. In other words, this is reminiscent of two different solids, initially kept at different temperatures and then brought together. Again, in this case, the two subsystems reach the same temperature but different chemical potentials since there is no power transfer.

Figure 4 shows the possibility of two different optical nonlinear multimoded subsystems (a Lieb and a square lattice) exchanging both energy and power, while the total energy $U_T = U_x + U_y$ and total power $\mathcal{P}_T = \mathcal{P}_x + \mathcal{P}_y$ are conserved. This grand canonical-like ensemble is analogous to that of two solids or gases that simultaneously allow both heat and particle transfer. A possible design for such an optically permeable wall that allows in addition, four-wave mixing mediated power exchange between the $\hat{x}$ & $\hat{y}$ polarizations, is provided in Supplementary Methods Fig. S3. As before, cross-phase modulation is responsible for energy exchange. Before merging together, the subsystems are at thermal equilibrium having $(T_x, \mu_x) = (-0.17, 4.21)$ and $(T_y, \mu_y) = (0.065, -4.956)$. Once in contact, computer simulations show that they now attain not only the same temperature $T = -0.42$ but also the same chemical potential $\mu = 6.46$, in excellent agreement with results anticipated from the theoretical formalism developed above. Again, the response of the combined system is driven by the second law of thermodynamics, since now $dS_T = (T_x^{-1} - T_y^{-1})dU_x + (\mu_y T_y^{-1} - \mu_x T_x^{-1})d\mathcal{P}_x \geq 0$. This latter equation directly implies that, in addition to having energy transfer from a hot to a cold object, the optical power will flow between the two polarizations in this multimode arrangement towards the subsystem with a lower chemical potential until thermal equilibrium is reached.

We now consider the interesting possibility of an all-optical Carnot-like cycle mediated by successive adiabatic or isentropic ($dS = 0$) expansions and compressions, taking place between a cold and a hot subsystem. An example of a nonlinear multimode array undergoing an adiabatic expansion is shown in Fig. 5a. During this process, the number of modes remains the same while the discrete Hamiltonian energy $U$ decreases during expansion as a result of the reduction in coupling strengths (see Methods). Similarly, during compression the energy increases. Under adiabatic conditions, the mode occupancies $|c_i|^2$ remain invariant and as a result the process is isentropic, given that $S = \sum \ln|c_i|^2$. From here, one can formally prove that the following laws hold during isentropic compressions or expansions:

$$\frac{\mu}{T} = \text{constant} \quad (5a)$$

$$\frac{U}{T} = \text{constant} \quad (5b)$$

Interestingly, Eq. 5a is fully analogous to that expected from an ideal monoatomic gas undergoing isentropic transformations. Meanwhile, Eq. 5b indicates that as the energy goes up or down, so does the temperature in an analogous fashion. In other words, during compression the temperature increases while it decreases during expansion. These effects will now be utilized so as to implement an all-optical "refrigeration" cycle whereby energy is extracted from a cold MM lattice (right) and transferred to a hotter system (left) – see Fig. 5b. The optical multimode system in between acts as a "refrigerant" whereby the temperature first falls after expansion, thus inducing energy extraction from the cold object upon contact. After it gets detached from the right arrangement, it undergoes a compression (thus increasing its temperature) in order to deliver the extracted energy to the hot system on the left. In essence, the three subsystems in Fig. 5b are allowed to exchange energy via contact layers but not optical power, as in the canonical-like ensemble arrangement presented in Fig. 3. The Carnot-like cycle corresponding to this process is depicted in Fig. 5c. Such optical refrigeration schemes can be judiciously deployed in high power multimode fiber sources where the aim is to eventually drive the optical power into the lower group of modes via cooling – thus producing a high quality output beam that is nearly free of speckle.

In conclusion, we have developed a new thermodynamical formalism that can be utilized in a versatile manner to not only explain but also predict the complex response of nonlinear heavily multimoded optical systems. The thermodynamic laws derived here are universal. This theoretical framework can be used to devise novel all-optical techniques through which the equilibrium modal distribution (temperature, chemical potential) in a particular subsystem can be controlled at will. The principles derived here are not exclusive to optical structures but can be employed in a straightforward manner in numerous other bosonic arrangements.

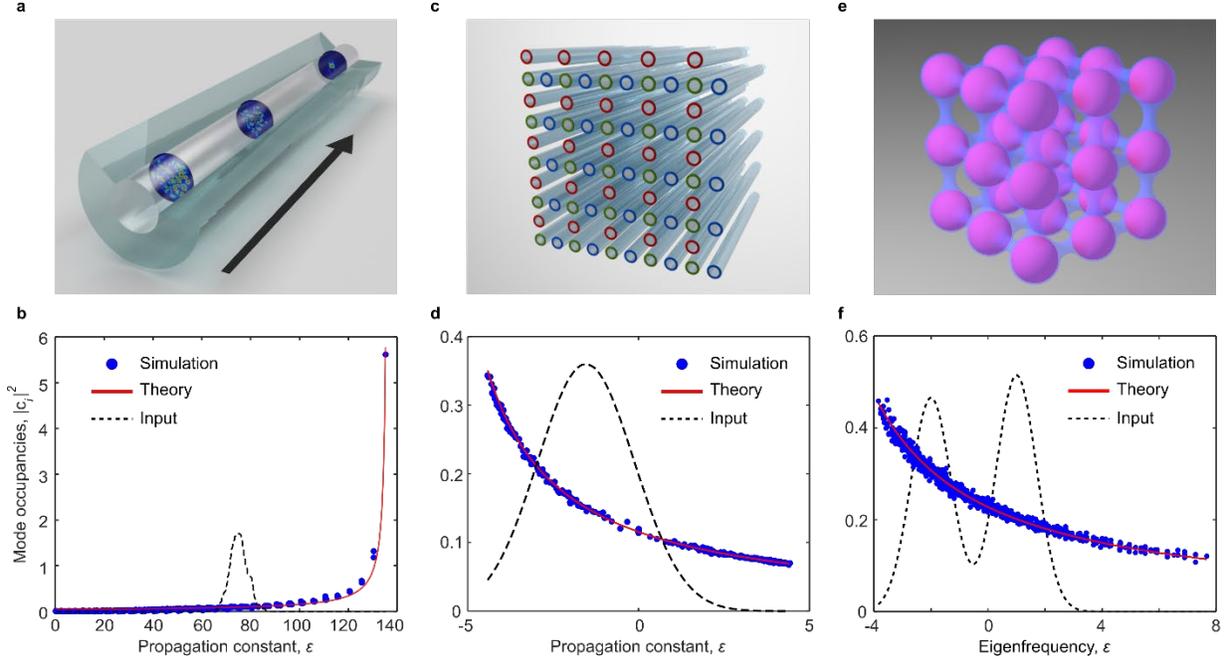

**Fig. 1 | Thermalization dynamics in nonlinear multimode waveguides and cavity structures. a**, A schematic of a graded-index multimode fiber system indicating the onset of thermalization, manifested as beam self-cleaning during propagation. **b**, The resulting mode occupancy (Rayleigh-Jeans) distributions after thermalization vs. normalized propagation constants after 30 m of propagation distance, as obtained from numerical simulations and theory when the parabolic-index fiber in **a** is excited with 180 kW at a wavelength of 532 nm. The dashed curve in **b** represents the modal excitation distribution at the input, with random initial phases. In this case, the final optical temperature is positive, $T = 5.37$, indicating a pronounced population in the lower-order modes. We note that in an actual experiment, thermalization can be attained at much smaller distances because of fiber perturbations that further promote ergodicity[46]. **c**, A nonlinear multicore Lieb waveguide lattice. **d**, Corresponding modal distributions at thermal equilibrium (for 300 modes as obtained from theory and simulations after propagating $\sim 10^4$ coupling lengths) for the initial excitation profile shown as a dashed curve, again with random initial phases. Here, the final temperature is negative, $T = -0.76$, thus promoting significant population in higher-order modes. The heavy degeneracy in the flat band of the Lieb lattice does not affect the thermalization process. **e**, A 3D coupled cavity nonlinear system. **f**, Resulting modal distributions after thermalization in time ($\sim 10^4$ normalized coupling cycles) in a configuration involving 1000 resonators, all cross-linked in a tight-binding model. The final temperature and chemical potential at equilibrium are $T = 1.75$ and $\mu = -7.7$, for the initial excitation conditions provided by the dashed curve. Note that in temporally evolving cavities, the spectrum is reversed since higher-order modes have lower eigenfrequencies.

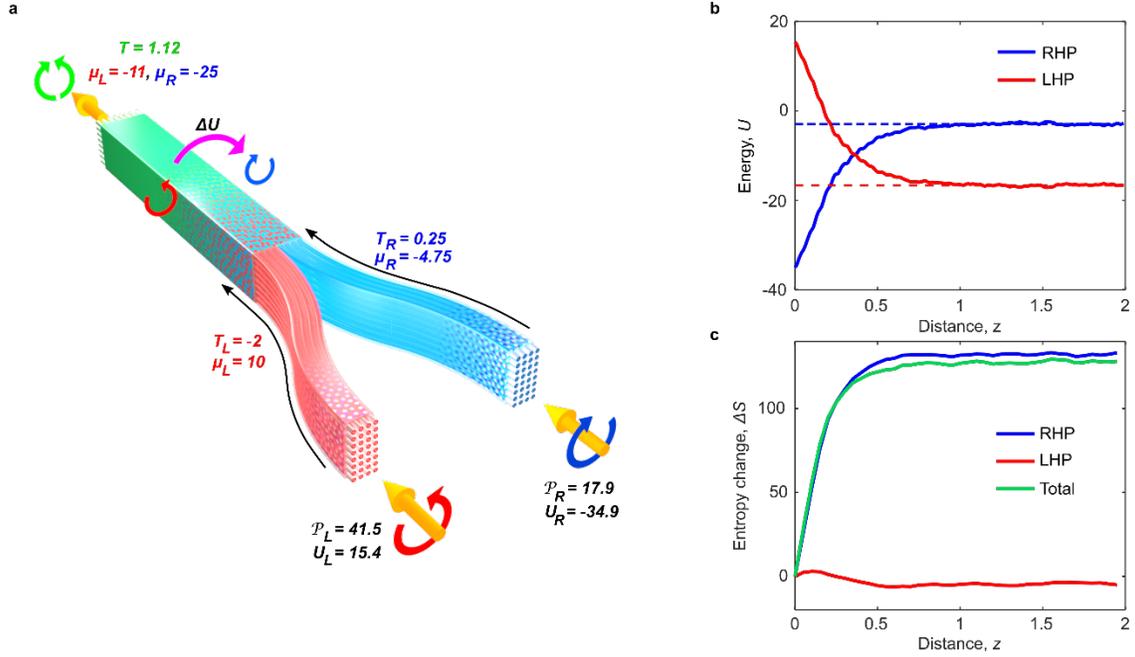

**Fig. 2 | Thermalization in a canonical-like optical multimode nonlinear configuration involving two circular polarizations**. **a**, A schematic of two separate waveguide arrays, each with 200 sites. The lattice on the left (L) is excited with a left-circular polarization (LHP) while the one on the right (R) with a right-handed (RHP). The corresponding initial power and internal energies are listed in the figure. Once the two subsystems reach thermal equilibrium, $(T_L, \mu_L) = (-2, 10)$ and $(T_R, \mu_R) = (0.25, -4.75)$, they merge together in a lattice structure having 400 modes. **b**. After merging, the two species (LHP and RHP) are now allowed to exchange energy $\Delta U$, from hot (red) to cold (blue), as also indicated in **a**, in full accord with the second law of thermodynamics. Eventually the two circular polarizations reach a common temperature $T = 1.12$, albeit with different chemical potentials, as predicted by theory and in agreement with numerical simulations. Dashed lines indicate the theoretical predictions for the thermalized energy in each of the two species. **c**. Corresponding changes in entropy for the two polarizations in the combined lattice, given that $S_{\text{Total}} = S_{\text{LHP}} + S_{\text{RHP}}$. In all cases, maximization of the entropy indicates optical thermal equilibrium.

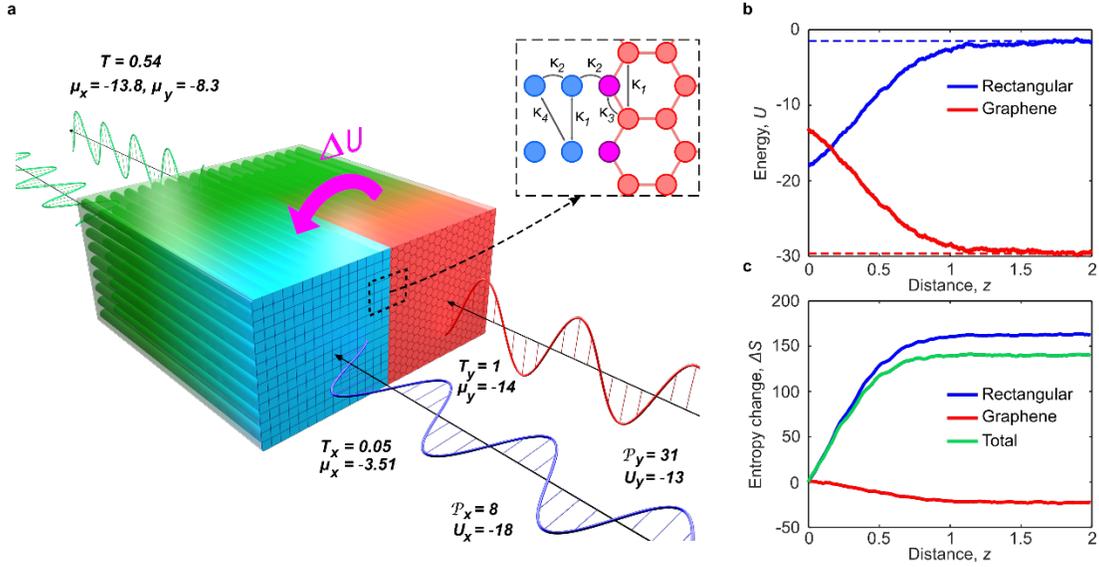

**Fig. 3 | Thermalization of two different optical waveguide lattices in thermal contact. a**, An optical graphene-like array (right), having 420 sites is brought in contact with a 200 site rectangular lattice (left) as shown in the inset, after the two lattices separately reach thermal equilibrium. The graphene lattice is excited with $\hat{y}$ polarized light and the rectangular one with $\hat{x}$. The thermal contact layer (shown in purple) allows the two solids to exchange energy $\Delta U$ via polarization cross-phase modulation while preventing any power transfer between the two lattices ($\mathcal{P}_x$ and $\mathcal{P}_y$ remain invariant). The two subsystems, set initially at $(T_y, \mu_y) = (1, -14)$ and $(T_x, \mu_x) = (0.05, -3.51)$, eventually reach a common temperature $T = 0.54$. Unlike in Fig. 2, where the two gases share the same vessel, here energy transfer only takes place through a single diathermic layer. **b**. Evolution of energy transfer as a function of distance. As before, dashed lines show the predictions for energy in each of the two species at equilibrium. **c**, Corresponding entropy changes in the two lattices where again $S_{\text{Total}} = S_x + S_y$.

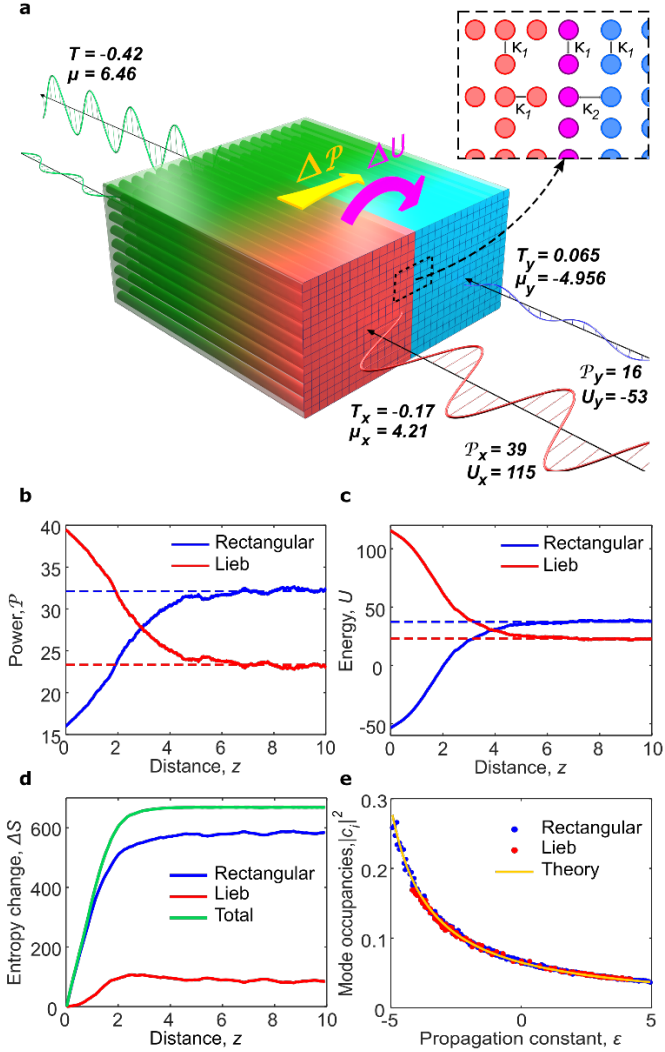

**Fig. 4 | Thermalization in a grand-canonical like configuration involving a Lieb and a rectangular nonlinear optical lattice. a**, As in Fig. 3, the two different arrays are brought together after each one of them reaches thermal equilibrium for the parameters provided in the figure. The Lieb subsystem on the left has 300 sites and is excited with $\hat{x}$ polarized light while the one on the right involves 400 modes and carries the $\hat{y}$ polarization. In addition to allowing energy transfer $\Delta U$, the contact layer is now designed in such a way so as to permit power exchange $\Delta \mathcal{P}$ between the two polarizations (via four-wave mixing). **b-c**, Evolution of optical power and internal energy during propagation. In this case, both the optical power and internal energy settle down to the predicted values (dashed lines). **d**, Corresponding entropic evolution for the entire system and the two subsystems. **e**, Resulting RJ distribution when the entire system finally thermalizes at the same negative temperature $T = -0.42$, as predicted by theory and confirmed numerically. In this case, the two species reach the same chemical potential as well.

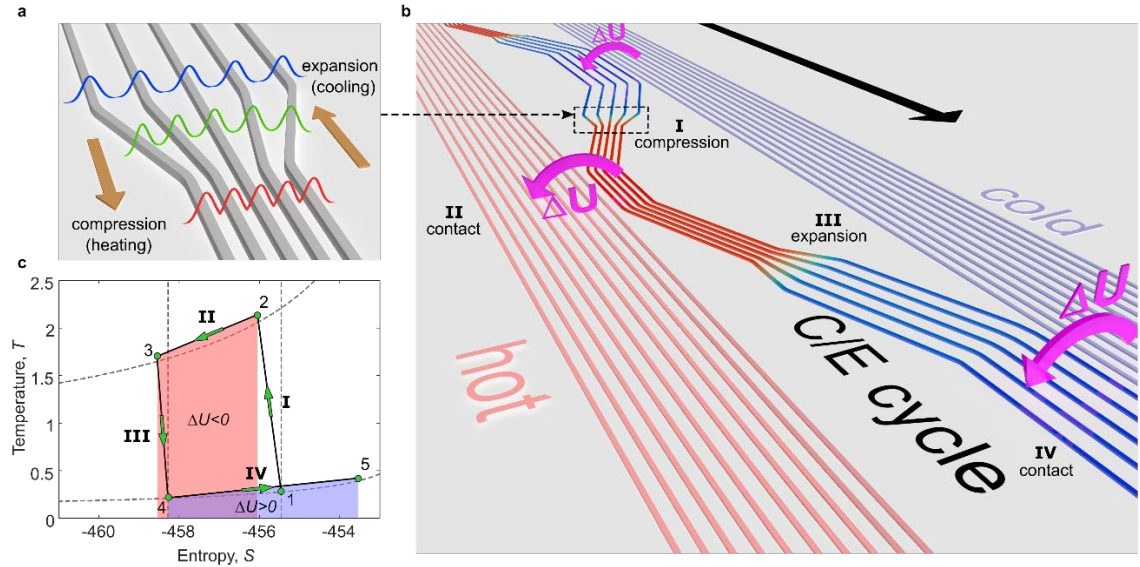

**Fig. 5 | All-optical refrigeration via Carnot cycles. a**, During adiabatic isentropic expansions, or compressions, the optical "gas" heats up or cools down, respectively. This is achieved by altering the internal energy $U$ through globally changing the coupling strengths between the individual elements comprising a multimode nonlinear optical array. **b**, Successive compression and expansion (C/E) cycles are used in order to achieve refrigeration in a particular subsystem (cold subsystem). In this configuration, three array subsystems are involved that are cyclically brought in thermal contact as shown in Fig. 3, allowing energy transfer $\Delta U$ from hot to cold. The array in the middle, acts as a refrigerant by undergoing a sequence of successive expansions and compressions. After expanding, its temperature drops below that of the cold optical array subsystem on the right, thus extracting energy from it and hence further cooling it down during contact. This excess energy is then passed on to the hotter subsystem on the left (during contact) after the array in the middle goes through an isentropic compression – during which its temperature exceeds that of the hot array. **c**, The temperature-entropy Carnot cycle corresponding to these four events. Dashed lines indicate the ideal Carnot cycle for this configuration.

# Methods

**Mode occupancy distribution**

We list the number of ways ($W$) in which one can distribute $N$ indistinguishable packets of power (or 'photons' at a specific wavelength) in $M$ distinct optical modes having energy levels $\varepsilon_i$ (propagation constants), each associated with a degeneracy $g_i$ ($g_i$ could also represent the number of clustered sub-levels[22] – see Extended Data Fig. 1). To do so, we assign to each $\varepsilon_i$ level group or cell, $n_i$ discrete power packets that are meant to be distributed over $g_i$ distinguishable compartments. From here one finds that $W_i = (n_i + g_i - 1)!/[n_i!\,(g_i - 1)!]$. Hence, the total number of ways $W$ can be obtained from,

$$W(n_1, n_2, n_3, \ldots) = \prod_i^M \frac{(n_i + g_i - 1)!}{n_i!\,(g_i - 1)!}$$

The entropy of the system is given by $S_N = \ln W$. Given that in actual settings $n_i \gg 1$, and that $\ln n! = n \ln n - n$ (Stirling approximation), we find,

$$S_N = \sum_i^M (n_i + g_i - 1)\ln(n_i + g_i - 1) - (n_i + g_i - 1) - n_i \ln n_i + n_i - \ln(g_i - 1)!.$$

The entropy must be then maximized under the following two constraints:

$$N = \sum_i^M n_i$$

$$E = -\sum_i^M n_i \varepsilon_i,$$

where $E$ represents the total "internal energy" in the system and $N$ the total number of power packets or "particles". Extremization by means of Lagrange multipliers leads to a Bose-Einstein distribution,

$$\frac{\partial}{\partial n_i}\left(S_N + \alpha \sum_i^M n_i + \beta \sum_i^M \varepsilon_i n_i\right) = 0$$

$$\frac{n_i}{g_i} = \frac{1}{e^{-\alpha - \beta \varepsilon_i} - 1}.$$

Keeping in mind that even in heavily multimoded nonlinear optical systems, the number of power (or energy) packets is much larger than the number of available modes in a cluster, $n_i \gg g_i$, we conclude that,

$$e^{-\alpha - \beta \varepsilon_i} \to 1,$$

$$e^{-\alpha - \beta \varepsilon_i} - 1 \simeq -\alpha - \beta \varepsilon_i,$$

In doing so, the Bose-Einstein distribution readily reduces to a Rayleigh-Jeans distribution, i.e.,

$$\frac{n_i}{g_i} = -\frac{1}{\alpha + \beta \varepsilon_i} \quad (6)$$

We would like to mention that in the optical multimode system, the eigenenergies are positioned according to $\varepsilon_1 \leq \varepsilon_2 \leq \varepsilon_3 \ldots \leq \varepsilon_M$, where $\varepsilon_M$ represents the energy of the lowest order mode (ground state) while $\varepsilon_1$ that of the highest. The reason for this particular choice of sequence is that we measure these propagation constants from the cladding region in which case the ground state has the highest eigenvalue. This situation is reversed in nonlinear MM cavity systems.

**Optical Entropy**

The optical entropy can be directly obtained from the equation:

$$S_N = \sum_i^M (n_i + g_i - 1) \ln(n_i + g_i - 1) - (n_i + g_i - 1) - n_i \ln n_i + n_i - \ln(g_i - 1)!$$

derived above. Given that in a nonlinear multimoded optical system, $n_i/g_i \gg 1$, i.e., each mode is very highly populated, one can then use a Taylor expansion and omit small terms:

$$S_N = \sum_i^M (n_i + g_i - 1) \ln\left[n_i\left(1 + \frac{g_i - 1}{n_i}\right)\right] - g_i + 1 - n_i \ln n_i - \ln(g_i - 1)!$$

$$= \sum_i^M (n_i + g_i - 1) \ln n_i + (n_i + g_i - 1)\left(\frac{g_i - 1}{n_i}\right) - g_i + 1 - n_i \ln n_i - \ln(g_i - 1)!$$

From here, since $n_i \gg g_i$, one quickly finds that $S_N \simeq \sum g_i \ln n_i$, or more conveniently the optical entropy can now be written as:

$$S_N = \sum_i^M \ln n_i ,$$

where in obtaining our last result we have set $g_i = 1$ since the degeneracies were just used for clustering purposes. Moreover, $M$ now represents the total number of modes. Interestingly, this same entropy also applies in other settings like for example image restoration[50]. In our optical system, the power in each mode $|c_i|^2$ is proportional to the number of infinitesimal power packets $n_i$ in each energy level. The proportionality factor in this case can be selected in such a way that $n_i = n_c |c_i|^2$, where $n_c$ represents the number of power packets per unit of normalized power. Hence, the entropy can now be viewed as the sum of two components,

$$S_N = \sum_{i=1}^M \ln(n_c) + \sum_{i=1}^M \ln|c_i|^2.$$

Since $n_c$ is a constant, the first part in the above equation represents a floor (a reference point) in the entropy while the second one denotes a more relevant entropic component that responds to nonlinear mode mixing. Hence from this point on, we write,

$$S = \sum_{i=1}^M \ln|c_i|^2. \quad (7)$$

By adopting the more conventional definitions for temperature and chemical potential, i.e., $\alpha = \mu(Tn_c)^{-1}$ and $\beta = (Tn_c)^{-1}$, Eq. (6) reduces to,

$$|c_i|^2 = -\frac{T}{\varepsilon_i + \mu}, \quad (8)$$

Finally, by again using $n_i = n_c|c_i|^2$, the optical observables associated with the total power $\mathcal{P}$ and internal energy $U$ (Supplementary) can be expressed as follows,

$$U = -\sum_i^M \varepsilon_i |c_i|^2 \quad (9)$$

$$\mathcal{P} = \sum_i^M |c_i|^2 \quad (10)$$

**Derivation of the first equation of state**

After manipulating the expressions for the optical internal energy and power, we obtain,

$$\frac{U}{T} - \frac{\mu}{T}\mathcal{P} = \sum_i^M \left[\left(\frac{\varepsilon_i}{\varepsilon_i + \mu}\right) - \left(\frac{-\mu}{\varepsilon_i + \mu}\right)\right] = \sum_i^M 1 = M$$

Hence,

$$U - \mu\mathcal{P} = MT. \quad (11)$$

**Predicting $T$ and $\mu$ of a microcanonical ensemble at equilibrium**

The temperature $T$ and chemical potential $\mu$ can be directly determined from either Eq. (9) or (10) by invoking the first equation of state Eq. (11) so as to eliminate one variable among $T$ and $\mu$. For example, using the equilibrium modal distribution in Eq. (8) and given that from Eq. (11), $\mu = \mathcal{P}^{-1}(U - MT)$, we find that,

$$\mathcal{P} = \sum_i^M -\frac{T}{\varepsilon_i + \mu} = \sum_i^M -\frac{T}{\varepsilon_i + \mathcal{P}^{-1}(U - MT)}.$$

For a given input power $\mathcal{P}$ and internal energy $U$, and once the eigenspectrum $\varepsilon_i$ for this $M$-mode system is known, the only unknown variable $T$ can be uniquely determined by solving the equation above. Note that the only acceptable solution for $T$ is the one that keeps each occupancy $|c_i|^2$ positive [Eq. (8)]. From here, one can then obtain the chemical potential $\mu$ through Eq. (11).

**Extensivity of the optical entropy**

As previously indicated, at equilibrium, the entropy of a given multimode optical system is written as,

$$S(U, M, \mathcal{P}) = \sum_i^M \ln|c_i|^2 = \sum_i^M \ln\left(-\frac{T}{\varepsilon_i + \mu}\right)$$

Let us now for example, double the optical arrangement $M \to 2M$ in such a way that the structure of the system, or the profile of the density of states, remains invariant (see Extended Data Fig. 2). At the same time, we double the input optical power and internal energy: $U \to 2U$, $\mathcal{P} \to 2\mathcal{P}$. From Eq. (11) we find that $T$ and $\mu$ still remain the same since $2U - 2\mu\mathcal{P} = 2MT$. The same conclusion can be reached by directly solving Eqs. (9) and (10).

Since each energy level $\varepsilon_i$ now splits into two closely spaced energy levels $\varepsilon_{i1}$ and $\varepsilon_{i2}$, where $\varepsilon_{i1} \approx \varepsilon_{i2} \approx \varepsilon_i$, the entropy now becomes,

$$S(2U, 2M, 2\mathcal{P}) = \sum_i^M \ln\left(-\frac{T}{\mu + \varepsilon_{i1}}\right) + \ln\left(-\frac{T}{\mu + \varepsilon_{i2}}\right)$$

Hence,

$$S(2U, 2M, 2\mathcal{P}) = 2\sum_i^M \ln\left(-\frac{T}{\mu + \varepsilon_i}\right) = 2S(U, M, \mathcal{P}).$$

In more general terms, $S(\lambda U, \lambda M, \lambda \mathcal{P}) = \lambda S(U, M, \mathcal{P})$

**Thermodynamic driving forces: temperature, chemical potential, and pressure**

We next show that the temperature can be self-consistently obtained from the fundamental equation of thermodynamics via:

$$\frac{\partial S}{\partial U} = \frac{1}{T} \quad (12)$$

Since $\mu = \mathcal{P}^{-1}(U - MT)$, from Eqs. (7) and (8) we obtain,

$$S = \sum_i^M \ln\left(\frac{-T}{\varepsilon_i + \mu}\right) = \sum_i^M \ln\left(\frac{T}{-\varepsilon_i - \frac{1}{\mathcal{P}}(U - MT)}\right)$$

$$\left.\frac{\partial S(U, M, \mathcal{P}; T)}{\partial U}\right|_{\mathcal{P},M} = \frac{\partial S}{\partial U} + \frac{\partial S}{\partial T}\frac{\partial T}{\partial U}$$

It can be directly seen that $\frac{\partial S}{\partial T} = 0$, since,

$$\frac{\partial S}{\partial T} = \sum_i^M \frac{1}{T} - \frac{M}{\mathcal{P}T}\sum_i^M \frac{T}{-\varepsilon_i - \mu} = \frac{M}{T} - \frac{M}{T} = 0.$$

This leaves us with,

$$\left.\frac{\partial S(U, M, \mathcal{P}; T)}{\partial U}\right|_{\mathcal{P},M} = \frac{\partial S}{\partial U} = \sum_i^M \frac{\frac{1}{\mathcal{P}}}{-\varepsilon_i - \frac{1}{\mathcal{P}}(U - MT)} = \frac{1}{T}\frac{1}{\mathcal{P}}\sum_i^M |c_i|^2$$

$$\left.\frac{\partial S}{\partial U}\right|_{\mathcal{P},M} = \frac{1}{T}$$

We next prove that the chemical potential can be formally defined according to,

$$\frac{\partial S}{\partial \mathcal{P}} = -\frac{\mu}{T}, \quad (13)$$

in a manner consistent with the first equation of state Eq. (11). If we instead write,

$$T = \frac{U - \mu \mathcal{P}}{M},$$

and substitute it into Eqs. (7) and (8), one obtains:

$$S = \sum_i^M \ln\left[\frac{U - \mu \mathcal{P}}{M(-\varepsilon_i - \mu)}\right]$$

$$\frac{\partial S(U, M, \mathcal{P}; \mu)}{\partial \mathcal{P}}\bigg|_{U,M} = \frac{\partial S}{\partial \mathcal{P}} + \frac{\partial S}{\partial \mu}\frac{\partial \mu}{\partial \mathcal{P}}$$

Similar to the last case, it can be directly shown that $\frac{\partial S}{\partial \mu} = 0$, since,

$$\frac{\partial S}{\partial \mu} = \sum_i^M \frac{-\mathcal{P}}{U - \mu \mathcal{P}} - \sum_i^M \frac{-M}{M(-\varepsilon_i - \mu)}$$

$$\frac{\partial S}{\partial \mu} = \frac{-M\mathcal{P}}{U - \mu \mathcal{P}} + \frac{1}{T}\sum_i^M \frac{T}{-\varepsilon_i - \mu}$$

$$\frac{\partial S}{\partial \mu} = -\frac{\mathcal{P}}{T} + \frac{\mathcal{P}}{T} = 0.$$

We are finally left with,

$$\frac{\partial S(U, M, \mathcal{P}, \mu)}{\partial P}\bigg|_{U,M} = \frac{\partial S}{\partial \mathcal{P}} = \sum_i^M \frac{-\mu}{U - \mu \mathcal{P}} = \frac{-\mu M}{U - \mu \mathcal{P}}$$

$$\frac{\partial S}{\partial \mathcal{P}}\bigg|_{U,M} = -\frac{\mu}{T}$$

We next define the third intensive variable $p$ ("optical pressure") through,

$$\frac{\partial S}{\partial M} = \frac{p}{T} \quad (14)$$

In essence, the variable $M$ plays the role of volume in conventional thermodynamics. As indicated in the text, any expansion in the number of modes $M$ should be carried out in a way that leaves the structure of the system the same, i.e., the profile of the density of states remains invariant (see Extended Data Fig. 2). In order to obtain the term $\partial S/\partial M$, we first define a density of states $D(\varepsilon) = dM/d\varepsilon$, providing the number of modes $D(\varepsilon)d\varepsilon$ per unit interval $d\varepsilon$. Hence, the total number of modes in the original system is given by $\int D(\varepsilon)d\varepsilon = M_0$. We now change in a self-similar manner the number of modes in this system in a way that preserves the profile of the density of states. In this case, the new density of states $\widetilde{D}(\varepsilon)$ can be obtained simply by: $\widetilde{D}(\varepsilon) = VD(\varepsilon)$, where $V$ is a scale factor indicating the fractional change in "volume" or the number of modes. From Eq. (8), we can rewrite the expressions for $U, \mathcal{P}$ and $S$ as follows,

$$U = \int \frac{\varepsilon T}{\varepsilon + \mu} V D(\varepsilon) d\varepsilon$$

$$\mathcal{P} = -\int \frac{T}{\varepsilon + \mu} V D(\varepsilon) d\varepsilon$$

$$S = \int \ln\left(-\frac{T}{\varepsilon + \mu}\right) V D(\varepsilon) d\varepsilon.$$

The first equation of state Eq. (11) leads to:

$$U - \mu\mathcal{P} = T \int V D(\varepsilon) d\varepsilon = V M_0 T = MT.$$

If we rewrite this equation as, $\mu = (U - VM_0 T)/\mathcal{P}$ and substitute into the expression for entropy, we get:

$$S = \int \ln\left[\frac{T}{-\varepsilon - \frac{1}{\mathcal{P}}(U - VM_0 T)}\right] V D(\varepsilon) d\varepsilon.$$

Upon differentiating with respect to $V$ one finds:

$$\left.\frac{\partial S(U, V, \mathcal{P}; T)}{\partial V}\right|_{\mathcal{P}, U} = \frac{\partial S}{\partial V} + \frac{\partial S}{\partial T}\frac{\partial T}{\partial V}$$

since $\partial S/\partial T = 0$ [as in the case of Eq. (12)]. From here one finds that:

$$\left.\frac{\partial S}{\partial V}\right|_{\mathcal{P}, U} = \frac{p}{T} = \frac{S}{V} - M_0.$$

By setting the arbitrary reference point $M_0$ to be 1, hence allowing $V \to M$, we then obtain,

$$\left.\frac{\partial S}{\partial M}\right|_{\mathcal{P}, U} = \frac{p}{T} = \frac{S}{M} - 1 \quad (15)$$

**Euler equation: second equation of state**

The extensive nature of entropy enables us to write:

$$S(\lambda U, \lambda M, \lambda \mathcal{P}) = \lambda S(U, M, \mathcal{P})$$

Differentiation on both sides of this last relation with respect to $\lambda$ yields an expression for $S$ in terms of all the variables involved in the multimoded system, better known as the Euler equation:

$$\frac{\partial S(\lambda U, \ldots)}{\partial (\lambda U)} U + \frac{\partial S(\lambda U, \ldots)}{\partial (\lambda M)} M + \frac{\partial S(\lambda U, \ldots)}{\partial (\lambda \mathcal{P})} \mathcal{P} = S(U, \mathcal{P}, M)$$

$$\frac{U}{T} + \frac{p}{T} M - \frac{\mu}{T} \mathcal{P} = S. \quad (16)$$

Where in deriving Eq. (16), we used Eqs. (12-14). The Euler equation can also be consistently obtained from Eq. (11) and Eq. (15). The extensivity of entropy with respect to $U$, $M$ and $\mathcal{P}$ is also apparent from Eq. (16).

**Direction of energy flow between two multimoded systems in thermal contact**

When two systems, both have initially reached thermal equilibrium independently, are allowed to exchange energy but not power (in a canonical-like ensemble), the second law of thermodynamics demands that the total entropy $S_T = S_1 + S_2$, should never decrease, i.e., $dS_T = dS_1 + dS_2 \geq 0$. Since, the power and volume associated with each subsystem is constant, the change in entropy depends only on energy exchanges, $dS = \frac{1}{T} dU$. Hence,

$$dS_T = \frac{1}{T_1} dU_1 + \frac{1}{T_2} dU_2 \geq 0$$

Since the total internal energy is conserved ($U_T = U_1 + U_2$ and $dU_T = 0$), $dU_1 = -dU_2$. Therefore,

$$\left(\frac{1}{T_1} - \frac{1}{T_2}\right) dU_1 \geq 0.$$

The last expression implies that the energy flows towards the first subsystem ($dU_1 > 0$) only when $(T_1^{-1} - T_2^{-1}) > 0$. In other words, if the temperatures of the two subsystems have the same sign, energy flows from the subsystem with a higher temperature to the one with the lower temperature; on the other hand, if the temperatures have a different sign, energy flows from the subsystem with a negative temperature to the subsystem with a positive temperature. The latter unconventional behavior is depicted in Fig. 2a in the main text.

**Direction of power and energy flow between two multimoded systems in diffusive contact**

When two systems, which have initially reached their thermal equilibrium independently, are allowed to exchange both energy and power (as in a grand canonical-like ensemble), the second law of thermodynamics now demands that,

$$dS_T = \frac{1}{T_1} dU_1 + \frac{1}{T_2} dU_2 - \frac{\mu_1}{T_1} d\mathcal{P}_1 - \frac{\mu_2}{T_2} d\mathcal{P}_2 \geq 0$$

Because of the two conservation laws: $dU_1 = -dU_2, d\mathcal{P}_1 = -d\mathcal{P}_2$. Thus,

$$\left(\frac{1}{T_1} - \frac{1}{T_2}\right) dU_1 + \left(\frac{\mu_2}{T_2} - \frac{\mu_1}{T_1}\right) d\mathcal{P}_1 \geq 0$$

As before, energy flows from a hotter system to a colder system. If the temperature is common to both subsystems ($T_1 = T_2$), power will always flow from the subsystem with a higher chemical potential to that with a lower chemical potential, if the temperature is positive, and vice versa if it is negative, always in such a way that $(\mu_2 - \mu_1) T^{-1} d\mathcal{P}_1 > 0$. Once equilibrium is reached and the chemical potentials reach the same value, exchange of power ceases.

**Graphical way of predicting the final temperature of canonical-like ensembles consisting of several subsystems**

Given a certain level of optical power $\mathcal{P}$, the $U(T)$ relation of any subsystem (associated with a set of propagation eigenvalues $\varepsilon_i$) can be obtained individually, as done in the microcanonical ensemble described earlier in the Methods. Consider a canonical system consisting of two subsystems A and B, each having its own $U(T)$ relation, $U_A(T_A)$ and $U_B(T_B)$. Since in a canonical ensemble, at equilibrium, all of the subsystems reach the same temperature through energy exchange, while the total energy of the combined system $U_0 = U_A + U_B$ is conserved, we have:

$$U_A(T) = U_0 - U_B(T).$$

From here, the common equilibrium temperature $T$ of the subsystems can be graphically obtained through finding the intersection of the curve $U_A(T)$ with $[U_0 - U_B(T)]$. Extended Data Fig. 3 illustrates this graphical technique, as used to predict the final temperature associated with simulations in Figs. 2 and 3.

**Isentropic invariants**

During an adiabatic process, the mode occupancies $|c_i|^2$ remain the same, making it isentropic since $S = \sum \ln|c_i|^2$. Meanwhile, during this same process, the eigenenergies $\varepsilon_i$ are expected to change through a common proportionality factor and so does the internal energy $U$. Consequently, the temperature and chemical potential are expected to adiabatically change as well. If initially, the system is at equilibrium, the RJ-distribution requires, $|c_i|^2 = -T_0/(\varepsilon_i + \mu_0)$, where $T_0$ and $\mu_0$ represent the initial temperature and chemical potential, respectively. We now assume that during the isentropic process, $\varepsilon_i \to \lambda \varepsilon_i$, $T_0 \to T_1$ and $\mu_0 \to \mu_1$, in a way that the RJ-distribution is respected, i.e., $|c_i|^2 = -T_1/(\lambda \varepsilon_i + \mu_1)$. Since the mode occupancies $|c_i|^2$ are invariant during the process, we find that,

$$\frac{\varepsilon_i}{T_0} + \frac{\mu_0}{T_0} = \frac{\lambda \varepsilon_i}{T_1} + \frac{\mu_1}{T_1}, \quad \text{for all possible } i$$

If we write $\mu/T = \alpha$, then for two different arbitrary eigenstates, $i = k, m$, we obtain,

$$\varepsilon_k \left( \frac{1}{T_0} - \frac{\lambda}{T_1} \right) = \Delta \alpha$$

$$\varepsilon_m \left( \frac{1}{T_0} - \frac{\lambda}{T_1} \right) = \Delta \alpha$$

where $\Delta \alpha = \alpha_1 - \alpha_0$. If we assume that $\Delta \alpha \neq 0$, the above equations imply that for any $k$ and $m$, $\varepsilon_k = \varepsilon_m$, which is by itself a contradiction. The only way to avoid this contradiction is to set $\Delta \alpha = 0$ which shows that during an isentropic process,

$$\frac{\mu}{T} = \text{constant.}$$

On the other hand, the above equations suggest that $T_1 = \lambda T_0$ and $\mu_1 = \lambda \mu_0$. In addition, since,

$$U_1 = -\sum_i^M \lambda \varepsilon_i |c_i|^2 = \lambda U_0,$$

we find that,

$$\frac{U}{T} = \text{constant.}$$

**Terminology**

Throughout this paper, we loosely use the term "canonical-like ensemble" in order to describe two different subsystems capable of exchanging internal energy $U$ but no optical power $\mathcal{P}$ through a diathermic wall. This does not necessarily mean that the subsystem is placed in contact with a thermal bath at constant temperature. Meanwhile, we use the term "grand canonical-like ensemble" to refer to the case where the two subsystems can also in addition exchange optical power via a diathermic permeable wall.

**Simulations**

Due to the chaotic nature of nonlinear interactions, the modal occupancies are at each point, fluctuating. Given that thermodynamics deals with the *averaged* values for these occupancies, it is important to describe how these averages are obtained in carrying out numerical simulations. In general, the averaging process can be conducted over ensembles or over the propagation distance (or time). Both approaches are expected to be in agreement as long as the system is ergodic. A combination of these two schemes can also be employed to reduce the computation time. In all the simulations carried out in this article, we used a window size of ~100 units of normalized propagation distance (or time) to track the evolution of $S$, and equilibrium is determined once the entropy is maximized. The averaged values of modal occupancies at equilibrium are then determined by considering the last 20% of the data. On the other hand, one can also carry out ensemble averaging by conducting a number of simulations with random initial phases while exciting the same modal group (that also happens to preserve the values of the extensive variables $U$ and $\mathcal{P}$). We would like to mention that at equilibrium, the intensive parameters $T$ and $\mu$ are universally determined only by the system's three invariants $U$, $M$ and $\mathcal{P}$ (or $E$ when considering cavities), regardless of the specific modal groups initially excited.

In Fig. 1**a-b**, we used a chaotic Pullen-Edmonds potential having a refractive index profile $n(x,y) = n_0[1 - \Delta(X^2 + Y^2 + X^2Y^2)/R_0^2]$ truncated at $n(X,Y) = n_{clad} = 1.446$, where $R_0 = 25\mu m, n_0 = 1.46$ and $\Delta = 0.01$. The simulation is conducted by solving 20 ensembles associated with the (2+1)-D scalar nonlinear Schrödinger equation (NLSE) (see Supplementary Information) having a Kerr nonlinear coefficient $n_2 = 3.2 \times 10^{-20} m^2/W$. On the other hand, the evolution of the optical field in the waveguide lattices discussed throughout the paper, is governed by a discrete NLSE (see Supplementary Information). In Fig. 1**c-d**, the Lieb lattice has normalized coupling coefficients $\kappa_x = 2, \kappa_y = 1$ with no detuning between sites and no cross-coupling along the diagonal. In Fig. 1**e-f**, the CROW system is a $10 \times 10 \times 10$ structure with a cubic unit cell. All couplings along the edges are $\kappa_1 = 1$ and along the diagonals of each face are $\kappa_2 = \sqrt{0.03}$. In this temporal case, the equilibrium modal occupancies follow instead the slightly modified RJ distribution $|c_i|^2 = T/(\varepsilon_i - \mu)$ in order to be consistent with the thermodynamic driving forces mentioned earlier. In Fig. 2, the coupling between all adjacent sites is $\kappa_1 = 1$ and on the diagonal it is $\kappa_2 = \sqrt{0.03}$. When the two lattices are brought together, the combined system retains the same geometric structure. In this case, the normalized nonlinear coefficient was $\gamma = 0.1$ and the normalized propagation distance is scaled in units of $10^4$ coupling lengths. The data presented corresponds to an ensemble average over 30 samples. In Fig. 3, the values of coupling strengths indicated in the inset are $\kappa_1 = 0.5\sqrt{4/5}, \kappa_2 = 1, \kappa_3 = \sqrt{4/5}, \kappa_4 = \sqrt{0.03}$ and 5 ensembles were used in the averaging process. Here, the normalized propagation distance is scaled in units of $10^5$ coupling lengths. The coupling values used in Fig. 4 are $\kappa_1 = 1.5$ and $\kappa_2 = 1$. The plotted data corresponds to an ensemble average over 20 samples. The distance axes are scaled in units of $10^4$ coupling lengths. In all of our simulations associated with discrete systems, the normalized power per mode ($\mathcal{P}/M$) is kept ~0.1 so as to allow the systems to evolve in a quasi-linear fashion.

Finally, the lattice subsystems in Fig. 5 are shown in 1-D to illustrate the concept, while in actual simulations we used a 2-D $10 \times 20$ square/rectangular structure for each lattice, with no diagonal-couplings. The hot and cold arrays have $\kappa_x = \kappa_y = 1$, while the coupling strengths in the C/E array are $(\kappa_x, \kappa_y) = \lambda(1, \sqrt{5/4})$ where the coupling factor $\lambda$ varies as $2 \rightarrow 0.25$ during expansion and vice versa during compression. The optical power in the left, middle and right arrays are $\mathcal{P} = 27, 21$ and $7.6$, respectively. The data points $1 - 5$ indicated by green dots were obtained when the array in the middle was detached from either of the two sublattices on the sides (see Extended Data Fig. 4). The dashed vertical lines in Fig. 5**c** represent the ideal isentropic processes. The upper and lower dashed curves were obtained by finding the $T - S$ curve of the array in the middle when $\lambda$ was set equal to 2 and 0.25, respectively.

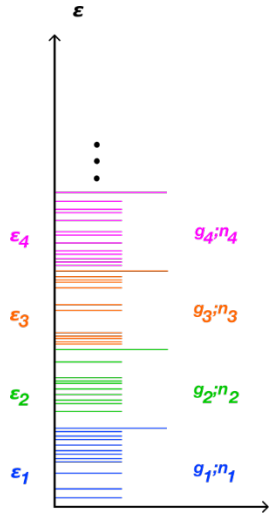

**Extended Data Fig. 1 | Clustering energy levels into groups.** The eigen-energy spectrum is determined by the optical system under consideration. On the other hand, the way of grouping the states in each cell with an associated degeneracy $g_i$, does not affect the final equilibrium distribution.

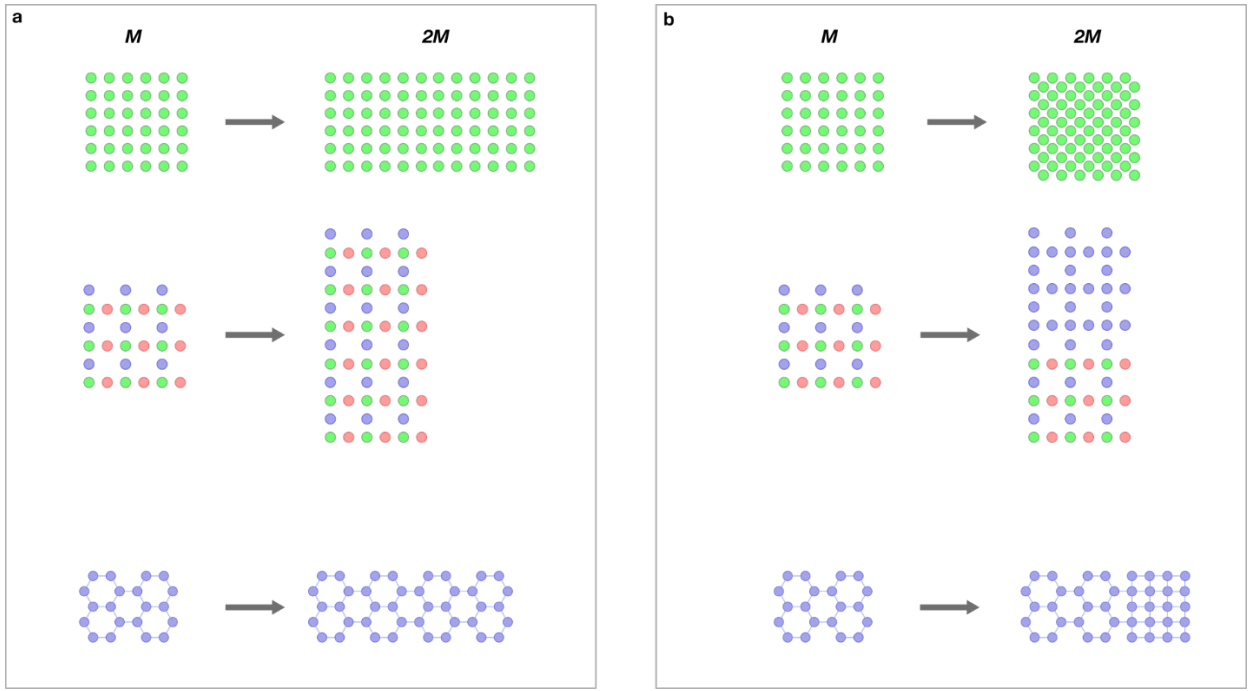

**Extended Data Fig. 2 | Different ways of enlarging the optical volume *M*. a**, Doubling the number of sites *M* in a self-similar way so as the profile of the density of states remains invariant. **b**, Doubling *M* in a way that does not respect self-similarity – the new material is now analogous to an alloy.

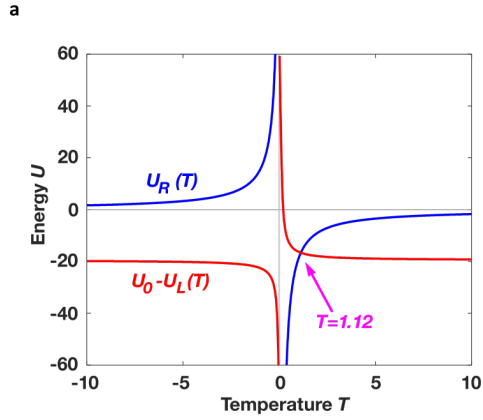 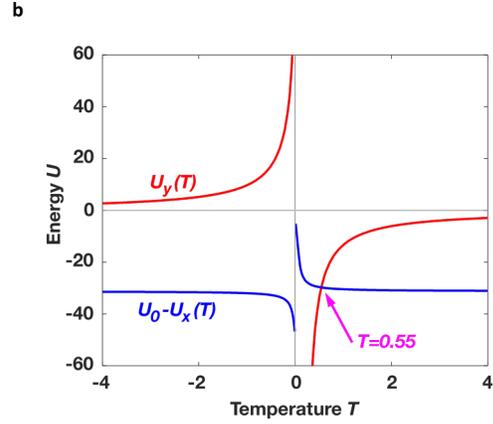

**Extended Data Fig. 3 | Graphically predicting the equilibrium temperature of a canonical-like ensemble. a**, $U_R(T)$ and $U_L(T)$ represent the energy-temperature curves associated with right-hand and left-hand circular polarizations, respectively, in Fig. 2 of the main text. The intersection of the two curves, $U_R(T)$ and $U_0 - U_L(T)$ provides the equilibrium temperature $T = 1.12$ of such a canonical-like ensemble. **b**, Similarly, the temperature is predicted for the combined system in Fig. 3, consisting of a square lattice excited with $\hat{x}$-polarized light and a graphene lattice excited with $\hat{y}$-polarized light.

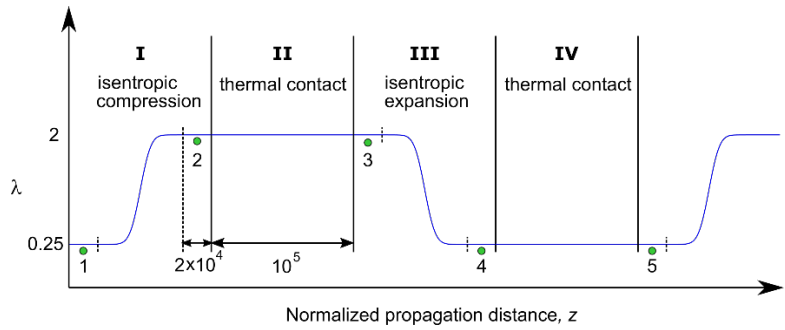

**Extended Data Fig. 4 | Coupling profile for the C/E cycles.** The variation of the coupling factor $\lambda$ for the middle array is shown as a function of propagation distance. Each of the four processes (**I-IV**) take place over a normalized propagation distance of $10^5$. In **I** and **III**, the coupling factor $\lambda$ is varied as an error function, and the temperature $T$ and entropy $S$ are found by averaging over the interval $2 \times 10^4$ at the locations denoted by $1 - 5$ (that also correspond to Fig. 5**c**).

## Supplementary Methods

**Normalizations and conserved quantities in a discrete optical system**

We here provide the normalizations carried out in a 1-D photonic lattice, as mentioned in the main text. In such a structure, evolution of the complex amplitude $A_m$ of the optical field at the $m^{th}$ site is governed by:

$$i\frac{dA_m}{dZ} + \beta_m A_m + \kappa(A_{m+1} + A_{m-1}) + n_2 k_0 |A_m|^2 A_m = 0,$$

where $\beta_m$ is the propagation constant of the corresponding site, $\kappa$ is the evanescent-wave coupling coefficient between neighboring sites, $k_0$ is the wavenumber in vacuum and $n_2$ represents the effective nonlinear coefficient.

If we let $z = \kappa Z$, $a_m = \frac{A_m}{\rho} e^{-i\beta_0 Z}$ ($\beta_0$ represents an arbitrary propagation constant that can serve as a reference, like for example the average value of all propagation constants) and divide both sides by $\rho e^{i\beta_0 Z}$, we get,

$$i\frac{da_m}{dz} + \frac{(\beta_m - \beta_0)}{\kappa} a_m + a_{m+1} + a_{m-1} + \frac{n_2 k_0 \rho^2}{\kappa} |a_m|^2 a_m = 0.$$

Upon setting $\Delta_m = \frac{(\beta_m - \beta_0)}{\kappa} \in \mathbb{R}$, and choosing a $\rho$ such that $\gamma = \frac{n_2 k_0 \rho^2}{\kappa} = 1$ (if the value of $\gamma$ is not specified, it is assumed to be unity throughout the paper), one obtains,

$$i\frac{da_m}{dz} + \Delta_m a_m + a_{m+1} + a_{m-1} + |a_m|^2 a_m = 0.$$

The linear part of the above equation can be written in a more compact form as follows:

$$i\frac{\partial |\Psi\rangle}{\partial z} + \widehat{H}|\Psi\rangle = 0,$$

where $|\Psi\rangle = (a_1, a_2, \ldots, a_M)^T$ is the state vector and $\widehat{H}$ denotes the Hamiltonian matrix operator of the system with eigenvalues $\epsilon_i$ and with corresponding orthonormal eigenvectors $|\phi_i\rangle$.

The full dynamical system Hamiltonian can be written as,

$$H(a_m, ia_m^*) = \sum_{m=1}^{M} \Delta_m |a_m|^2 + a_m a_{m+1}^* + a_m^* a_{m+1} + \frac{1}{2} |a_m|^4,$$

which can be split into a linear and a nonlinear component,

$$H_L = \sum_{m=1}^{M} \Delta_m |a_m|^2 + a_m a_{m+1}^* + a_m^* a_{m+1},$$

$$H_{NL} = \sum_{m=1}^{M} \frac{1}{2} |a_m|^4.$$

The equation of motion can be obtained from the canonical equations of Hamilton:

$$\frac{\partial H}{\partial a_m} = \frac{\partial i a_m^*}{\partial z},$$

$$\frac{\partial H}{\partial i a_m^*} = -\frac{\partial a_m}{\partial z}.$$

Using the equations above:

$$\frac{\partial H}{\partial a_m} = \Delta_m a_m^* + a_{m+1}^* + a_{m-1}^* + |a_m|^2 a_m^* = \frac{\partial i a_m^*}{\partial z},$$

$$\frac{\partial H}{\partial i a_m^*} = -i(\Delta_m a_m + a_{m+1} + a_{m-1} + |a_m|^2 a_m) = -\frac{\partial a_m}{\partial z}.$$

These two equations are equivalent to the aforementioned evolution equation. The Hamiltonian is conserved during evolution since it does not explicitly depend on z, i.e., $\partial H/\partial z = 0$:

$$\frac{dH(a_m, i a_m^*)}{dz} = \frac{\partial H(a_m, i a_m^*)}{\partial a_m}\frac{\partial a_m}{\partial z} + \frac{\partial H(a_m, i a_m^*)}{\partial i a_m^*}\frac{\partial i a_m^*}{\partial z} + \frac{\partial H(a_m, i a_m^*)}{\partial z}$$

$$= \frac{\partial i a_m^*}{\partial z}\frac{\partial a_m}{\partial z} - \frac{\partial a_m}{\partial z}\frac{\partial i a_m^*}{\partial z} + \frac{\partial H(a_m, i a_m^*)}{\partial z}$$

$$= \frac{\partial H(a_m, i a_m^*)}{\partial z} = 0.$$

In addition, the same holds for the optical power, $\mathcal{P} = \sum_{i=1}^{M}|c_i|^2 = \sum_{m=1}^{M}|a_m|^2$, since

$$\frac{d\mathcal{P}}{dz} = \sum_{i=1}^{M} a_m^* \frac{\partial a_m}{\partial z} + a_m \frac{\partial a_m^*}{\partial z} = 0.$$

In a Hermitian system, the linear part of the Hamiltonian can be written in a modal representation using the fact that the eigenmodes are orthonormal, i.e.,

$$H_L = \langle \Psi|\hat{H}|\Psi\rangle = \sum_i^M c_i^* \langle \phi_i|\hat{H}\sum_k^M c_k|\phi_k\rangle = \sum_i^M \varepsilon_i |c_i|^2.$$

where $c_i = \langle \phi_i|\Psi\rangle$ are the mode occupancies. Since $H_L \simeq H$ is conserved, the internal energy ($U = -H_L$) in an optical system can be uniquely determined in terms of the initially excited mode group $|c_{i0}|^2$ once the input state vector $|\Psi_{in}\rangle$ is specified, i.e.,

$$U = -H_L = -\langle \Psi_{in}|\hat{H}|\Psi_{in}\rangle = -\sum_i^M \varepsilon_i |c_{i0}|^2.$$

**Optical field normalizations in a nonlinear optical fiber with an arbitrary index profile**

In the paraxial regime, the evolution of the optical field $E$, confined in a refractive index potential $n(X,Y)$, follows a Schrödinger-type equation,

$$\frac{i\partial E}{\partial Z} + \frac{1}{2k}\frac{\partial^2 E}{\partial X^2} + \frac{1}{2k}\frac{\partial^2 E}{\partial Y^2} + n(X,Y)E + n_2 k_0 |E|^2 E = 0,$$

where $k = k_0 n_0$ is the wavenumber and $n_2$ is the nonlinear coefficient. By using $Z = 2k_0 n_0 x_0^2 z, X = xx_0, Y = yx_0, E = \rho\psi e^{ik_0 n_{clad} Z}$ ($n_{clad}$ is the refractive index of the fiber cladding) and $\rho = 1/\sqrt{2k_0^2 n_0 x_0^2 n_2}$, where $x_0$ is an arbitrary length scale for normalization, we obtain

$$i\frac{\partial \psi}{\partial z} + \frac{\partial^2 \psi}{\partial x^2} + \frac{\partial^2 \psi}{\partial y^2} + V(x,y)\psi + |\psi|^2 \psi = 0,$$

where $V(x,y) = 2k_0^2 n_0 x_0^2 [n(x,y) - n_{clad}]$. Under linear conditions (weak nonlinearity), the above equation can be written as

$$i\frac{\partial \psi}{\partial z} + \nabla_\perp^2 \psi + V(x,y)\psi = 0.$$

The eigenstates (modes) $\Phi = \phi_i e^{i\varepsilon_i z}$ can be found from

$$-\varepsilon_i \phi_i + \nabla_\perp^2 \phi_i + V(x,y)\phi_i = 0 \quad - (S1)$$

$(S1) \times \phi_k^*$:

$$-\varepsilon_i \phi_i \phi_k^* + \phi_k^* \nabla_\perp^2 \phi_i + V(x,y)\phi_i \phi_k^* = 0 \quad - (S2)$$

If we now integrate Eq. ($S2$) over transverse coordinates,

$$\iint_{-\infty}^{+\infty} (-\varepsilon_i \phi_i \phi_k^* + \phi_k^* \nabla_\perp^2 \phi_i + V(x,y)\phi_i \phi_k^*) dx dy = 0$$

$$\iint_{-\infty}^{+\infty} \phi_k^* \nabla_\perp^2 \phi_i dx dy + \iint_{-\infty}^{+\infty} V(x,y)\phi_i \phi_k^* dx dy = \iint_{-\infty}^{+\infty} \varepsilon_i \phi_i \phi_k^* dx dy \quad - (S3)$$

The integral of the first term can be simplified using:

$$\iint_{-\infty}^{+\infty} \phi_k^* \nabla_\perp^2 \phi_i dx dy = \iint_{-\infty}^{+\infty} \phi_k^* \frac{\partial^2 \phi_i}{\partial x^2} dx dy + \iint_{-\infty}^{+\infty} \phi_k^* \frac{\partial^2 \phi_i}{\partial y^2} dy dx$$

$$= \int_{-\infty}^{+\infty} \left( \phi_k^* \frac{\partial \phi_i}{\partial x} \bigg|_{-\infty}^{+\infty} - \int_{-\infty}^{+\infty} \frac{\partial \phi_k^*}{\partial x} \frac{\partial \phi_i}{\partial x} dx \right) dy + \int_{-\infty}^{+\infty} \left( \phi_k^* \frac{\partial \phi_i}{\partial y} \bigg|_{-\infty}^{+\infty} - \int_{-\infty}^{+\infty} \frac{\partial \phi_k^*}{\partial y} \frac{\partial \phi_i}{\partial y} dy \right) dx,$$

since the eigenfunctions are bound states, i.e., $\phi_k^*(+\infty, -\infty) = 0$, hence,

$$\iint_{-\infty}^{+\infty} \phi_k^* \nabla_\perp^2 \phi_i dx dy = - \iint_{-\infty}^{+\infty} \left( \frac{\partial \phi_k^*}{\partial x} \frac{\partial \phi_i}{\partial x} + \frac{\partial \phi_k^*}{\partial y} \frac{\partial \phi_i}{\partial y} \right) dx dy \quad - (S4)$$

Using $\iint_{-\infty}^{+\infty} \phi_i \phi_k^* dx dy = \delta_{ik}$ where $\delta_{ik}$ is the Kronecker delta function, we obtain after substituting Eq. ($S4$) into Eq. ($S3$):

$$\iint_{-\infty}^{+\infty} \left( -\frac{\partial \phi_k^*}{\partial x} \frac{\partial \phi_i}{\partial x} - \frac{\partial \phi_k^*}{\partial y} \frac{\partial \phi_i}{\partial y} + V(x,y)\phi_i \phi_k^* \right) dx dy = \begin{cases} \varepsilon_i & (i = k) \\ 0 & (i \neq k) \end{cases} \quad - (S5)$$

An arbitrary optical field $\psi$ can, in general, be written as a superposition of eigenstates in the system

$$\psi(x,y,z) = \sum_{i=1}^{M} c_i \phi_i(x,y) e^{i\varepsilon_i z} \quad - (S6)$$

The averaged value of energy in such a continuous system, as opposed to the discrete case discussed before, can be written as

$$\langle H \rangle = \iint_{-\infty}^{+\infty} (-|\psi_x|^2 - |\psi_y|^2 + V(x,y)|\psi|^2) dx dy \quad - (S7)$$

Upon substituting Eq. (S6) into Eq. (S7), one finds,

$$\langle H \rangle = \sum_{i=1}^{M} \iint_{-\infty}^{+\infty} \left( -|c_i|^2 \left|\frac{\partial \phi_i}{\partial x}\right|^2 - |c_i|^2 \left|\frac{\partial \phi_i}{\partial y}\right|^2 + V(x,y)|c_i|^2|\phi_i|^2 \right) dx dy$$

$$+ \sum_{i=1}^{M} \sum_{k=1, k\neq i}^{M} c_i c_k^* e^{i(\varepsilon_i - \varepsilon_k)z} \iint_{-\infty}^{+\infty} \left( -\frac{\partial \phi_k^*}{\partial x}\frac{\partial \phi_i}{\partial x} - \frac{\partial \phi_k^*}{\partial y}\frac{\partial \phi_i}{\partial y} + V(x,y)\phi_i \phi_k^* \right) dx dy.$$

Hence, according to Eq. (S5),

$$U = -\langle H \rangle = \sum_{i=1}^{M} -\varepsilon_i |c_i|^2$$

**Lieb lattice: Governing equations**

The discrete coupled nonlinear Schrödinger equations describing the evolution of the normalized optical field at sites $a, b$ and $c$ (Supplementary Methods Fig. S1) in the Lieb lattice shown in Figs. 1**c-d** of the main text are given by

$$i\frac{da_{m,n}}{dz} + \kappa_1(b_{m,n} + b_{m,n+1}) = 0$$
$$i\frac{db_{m,n}}{dz} + \kappa_1(a_{m,n} + a_{m-1,n}) + \kappa_2(c_{m-1,n} + c_{m,n}) = 0$$
$$i\frac{dc_{m,n}}{dz} + \kappa_2(b_{m,n} + b_{m+1,n}) = 0$$

where $m, n$ denote the unit cell numbers, $\kappa_1$ and $\kappa_2$ are the vertical and horizontal nearest-neighbor coupling coefficients, respectively.

**3D array of optical cavities: governing equation**

The evolution of the normalized optical field $a_{l,m,n}$ in coupled resonator optical waveguides (CROWs) having a cubic unit cell (Fig. 1**e-f**) are described by

$$i\frac{da_{l,m,n}}{dt} + \kappa_1(a_{l-1,m,n} + a_{l+1,m,n} + a_{l,m-1,n} + a_{l,m+1,n} + a_{l,m,n-1} + a_{l,m,n+1})$$
$$+ \kappa_2(a_{l-1,m-1,n} + a_{l-1,m+1,n} + a_{l+1,m-1,n} + a_{l+1,m+1,n} + a_{l-1,m,n-1} + a_{l-1,m,n+1}$$
$$+ a_{l+1,m,n-1} + a_{l+1,m,n+1} + a_{l,m-1,n-1} + a_{l,m-1,n+1} + a_{l,m+1,n-1} + a_{l,m+1,n+1})$$
$$+ |a_{l,m,n}|^2 a_{l,m,n} = 0$$

where $l, m, n$ denote the site numbers along the $x, y, z$ coordinates, $\kappa_1$ and $\kappa_2$ are the coupling coefficients between cavities along the edges and along the diagonals in each unit cell, respectively. The last term in the equation represents the Kerr nonlinearity.

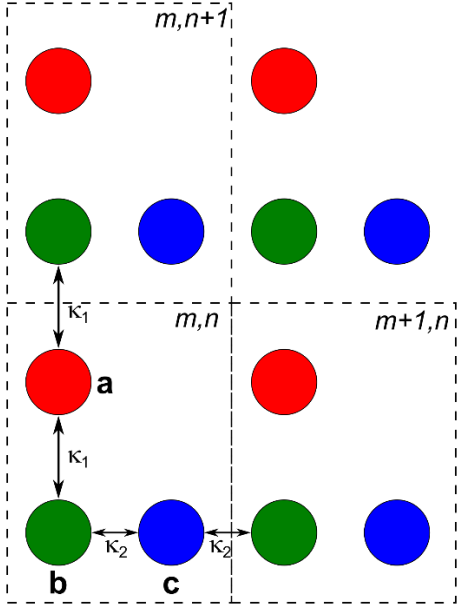

**Supplementary Methods Fig. S1 | Geometric structure of Lieb lattice.** Each unit cell in the Lieb lattice (dashed line) contains three sites, indicated by red, green and blue. When only the vertical and horizontal nearest neighbor couplings are taken into account, the Lieb lattice contains a highly degenerate flat band at the center of the eigen-spectrum.

**Governing equation for the two-species canonical-like and grand canonical-like ensembles**

Following the same normalizations as indicated earlier, the discrete coupled nonlinear Schrödinger equations describing the evolution of the normalized optical field for two orthogonal polarizations $a$ and $b$ in an optical lattice are given by

$$i\frac{da_m}{dz} + n_m^a a_m + \sum_{n=1,n\neq m}^{M} \kappa_{mn}^a a_n + A|a_m|^2 a_m + B|b_m|^2 a_m + C(b_m)^2 a_m^* = 0 \quad - (S8)$$

$$i\frac{db_m}{dz} + n_m^b b_m + \sum_{n=1,n\neq m}^{M} \kappa_{mn}^b b_n + A|b_m|^2 b_m + B|a_m|^2 b_m + C(a_m)^2 b_m^* = 0 \quad - (S9)$$

where $m$ denotes the site number, $\kappa_{mn}^{a(b)}$ are the coupling coefficients between site $m$ and site $n$, and the last three terms correspond to self-phase modulation, cross-phase modulation and four-wave mixing effects associated with the nonlinear coefficients $A, B$ and $C$, respectively. The Hamiltonian for this coupled system is conserved and can be written as:

$$H = \sum_{m}^{M} \sum_{n \neq m}^{M} \left( \kappa_{mn}^a a_m^* a_n + \kappa_{mn}^b b_m^* b_n \right) + \sum_{m}^{M} \left( n_m^a |a_m|^2 + n_m^b |b_m|^2 \right)$$
$$+ \sum_{m}^{M} \frac{1}{2} A(|a_m|^4 + |b_m|^4) + B|a_m|^2 |b_m|^2 + \frac{1}{2} C \left( a_m^{*2} b_m^2 + a_m^2 b_m^{*2} \right)$$

The coefficients $A, B$ and $C$ are determined by the optical structure and the polarization basis under consideration. In a canonical-like ensemble based on circular polarization in silica (Fig. 2), $A = 1, B = 2, C = 0$, while in Fig. 3 of the main text, for linear polarizations, $A = 1, B = 2/3, C = 0$. In both cases, the four-wave mixing term is negligible ($C = 0$), and one can easily infer that the optical powers in the two species $\mathcal{P}_a$ and $\mathcal{P}_b$ are individually conserved, as is required in a canonical-like ensemble. However, for the grand canonical-like case, as in Fig. 4, $A = 1, B = 2/3, C = 1/3$.

**Lattice design for a canonical-like setting**

We here present an optical lattice design that can act as a canonical-like ensemble, and where the field evolution is governed by Eqs. (S8) and (S9) when $A = 1, B = 2/3, C = 0$.

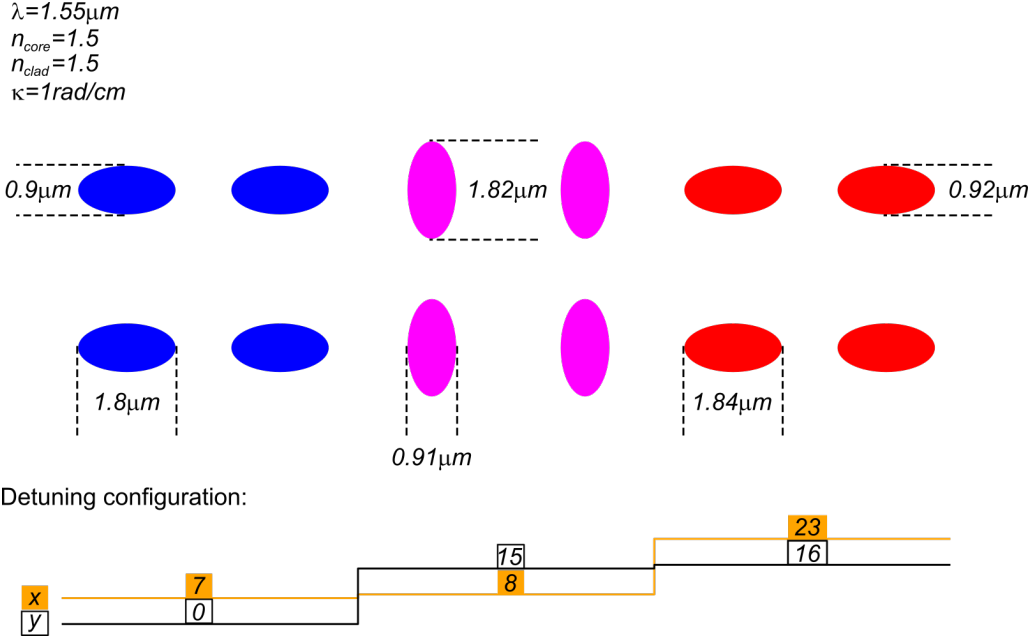

**Supplementary Methods Fig. S2 | Possible design for implementing a canonical-like setting.** The design contains three groups of polarization maintaining sites, based on their different geometries. The $\hat{x}$ and $\hat{y}$ polarizations are confined within the left (blue) and right (red) sides respectively but are allowed to overlap only in the thermal layer (purple) in between. The propagation constants for these two polarizations have been selectively tuned in order to achieve this. For example, for the $\hat{x}$ polarization, there is a negligible (propagation constant-) detuning between the blue and purple sites but a huge detuning between the right two groups, i.e., the purple and red sites. On the other hand, for the $\hat{y}$ polarization, there is a negligible detuning between the red and purple sites but a huge detuning between the left two groups. Therefore, the $\hat{x}$ polarization can travel freely in the left two groups without coupling into the right group (see the yellow line beneath the figure). Similarly, the $\hat{y}$ polarization stays confined in the right two groups (see the black curve beneath the figure). Moreover, to prevent any four-wave mixing mediated power exchange between these two polarizations ($C = 0$), each individual site is designed (elliptically) such that there exists a

substantial detuning between the $\hat{x}$ and $\hat{y}$ polarizations. In such a canonical-like multimoded optical setting, the left and right groups act as the two subsystems while the middle group plays the role of a thermal interaction layer.

**Lattice design for a grand canonical-like setting**

The design that can be treated as a grand canonical-like ensemble, and where the field evolution is governed by Eqs. (S8) and (S9) when $A = 1, B = 2/3, C = 1/3$ is present in the figure below.

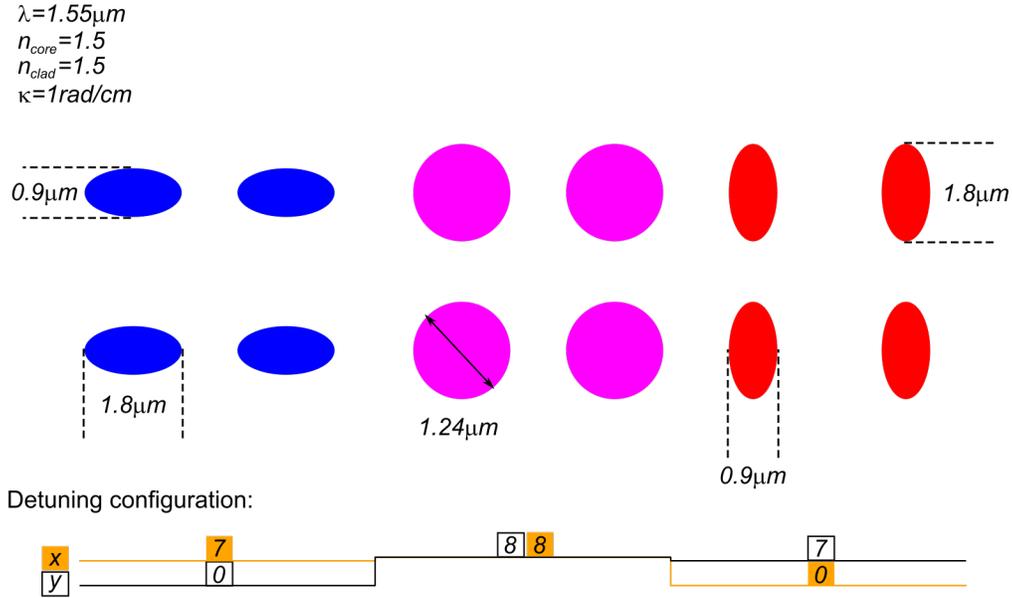

**Supplementary Methods Fig. S3 | Possible design for implementing a grand canonical-like setting.** The design contains two groups of polarization maintaining sites on the sides, and a group of circular waveguides in the middle acting as the interaction layer. Similar to the canonical case, the $\hat{x}$ and $\hat{y}$ polarizations are confined within the left (blue) and right (red) sides respectively and are allowed to overlap only in the thermal layer (purple) in between. However, in this case, the interaction layer does not provide enough birefringence to prevent the nonlinear four-wave mixing effect, i.e., $C \neq 0$. Therefore, optical power in the $\hat{x}$ polarization can be converted into $\hat{y}$ polarization via four-wave mixing and hence can be transferred to the right side. In the same vein, power in the $\hat{y}$-polarized optical field can flow to the left side. Hence, such a design can be treated as a grand canonical-like ensemble.